%% file: paper.tex
\documentclass[aps,prl,twocolumn,showpacs,superscriptaddress]{revtex4}  
\usepackage{graphicx}  
\usepackage{dcolumn}   
\usepackage{bm}        
\usepackage{amssymb}   

\input babarsym

\newcommand{\BABARPubYear}    {08}
\newcommand{\BABARPubNumber}  {028}
\newcommand{\SLACPubNumber} {13351}

\begin{document}

\leftline{\babar-PUB-\BABARPubYear/\BABARPubNumber }
\rightline{SLAC-PUB-\SLACPubNumber }

\title{Measurement of the Branching Fractions of {\boldmath $\Bbar \to D^{**} \ell^- \bar{\nu}_{\ell}$ } Decays in Events Tagged by a Fully Reconstructed {\boldmath $B$} Meson}

\input authors_jun2008.tex
\date{\today}

\begin{abstract}
We report a measurement of the branching fractions of $\Bbar \to D^{**} \ell^- \bar{\nu}_{\ell}$ decays based on 
417 fb$^{-1}$ of data collected at the $\Upsilon(4S)$ resonance with the \babar\
detector at the \pep2\ $e^+e^-$ storage rings. 
Events are selected by fully reconstructing one of the $B$ mesons in a hadronic decay mode.
A fit to the invariant mass differences
$m(D^{(*)}\pi)-m(D^{(*)})$ is performed to extract the signal yields of the different $D^{**}$ states.
We observe the $\Bbar \to D^{**} \ell^- \bar{\nu}_{\ell}$ decay modes corresponding to the four $D^{**}$ states predicted by  Heavy Quark Symmetry with a significance greater than six standard deviations including 
systematic uncertainties. 
\end{abstract}

\maketitle 

Semileptonic $B$ decays to orbitally-excited P-wave charm mesons ($D^{**}$) are of interest for several reasons. 
Improved knowledge of the branching fractions for these decays is important to reduce the systematic uncertainty in the measurements of the Cabibbo-Kobayashi-Maskawa~\cite{CKM} matrix elements $|V_{cb}|$ and $|V_{ub}|$. For example, one of the leading sources of
systematic uncertainty on $|V_{cb}|$ measurements from 
$\Bbar \to D^* \ell^- \bar{\nu}_{\ell}$ decays~\cite{ell} is the limited knowledge of the
background due to $\Bbar \to D^{**} \ell^- \bar{\nu}_{\ell}$~\cite{BaBarHQET}. 

The $D^{**}$ mesons contain one charm quark and one light quark with relative angular momentum $L=1$. According to Heavy Quark Symmetry (HQS)~\cite{IW}, they form one doublet of states with angular momentum $j \equiv s_q + L= 3/2$  $\left[D_1(2420), D_2^*(2460)\right]$ and another doublet with $j=1/2$ $\left[D^*_0(2400), D_1'(2430)\right]$, where $s_q$ is the light quark spin. Parity and angular momentum conservation constrain the decays allowed for each state. The $D_1$ and $D_2^*$ states decay through a D-wave to $D^*\pi$ and $D^{(*)}\pi$, respectively, and have small decay widths, while the $D_0^*$ and $D_1'$ states decay through an S-wave to $D\pi$ and $D^*\pi$ and are very broad. 

 $\Bbar \rightarrow D^{**} \ell^- \bar{\nu}_{\ell}$ decays constitute a significant fraction of $B$
 semileptonic decays~\cite{pdg} and may help to explain the discrepancy between the inclusive $\Bbar \to X\ell^- \bar{\nu}_{\ell}$   rate and the sum of the measured exclusive decay rates~\cite{pdg,babar-2,babar-3}.
The measured decay properties for $\Bbar \rightarrow D^{**} \ell^- \bar{\nu}_{\ell}$ can be compared with  the predictions of the Heavy Quark Effective Theory (HQET)~\cite{LLSW}. QCD sum rules~\cite{uraltsev} imply the strong dominance of $B$ decays to the narrow $D^{**}$ states over those to the wide ones, while some experimental data show the opposite trend~\cite{belle,delphi2005}. 

In this letter, we present the observation of $B$ semileptonic decays into the four excited $D$ mesons predicted by HQS and  measure the ${\cal B}(\Bbar \to D^{**} \ell^- \bar{\nu}_{\ell})$ branching fractions. 
The analysis is based on data collected with the \babar\ detector~\cite{detector} at the 
\pep2\ asymmetric-energy $e^+e^-$ storage rings at SLAC. The data consist of a total 
of 417~fb$^{-1}$ recorded at the $\Upsilon(4S)$ 
resonance, corresponding to approximately 460 million \BB\ pairs. An additional 40~fb$^{-1}$, taken at a center-of-mass (CM) energy 40 MeV below the $\Upsilon(4S)$ resonance, is used to study background from $e^+e^- \to f\bar{f}~(f=u,d,s,c,\tau)$ continuum events. A detailed GEANT4-based Monte Carlo (MC) simulation~\cite{Geant} of \BB\ and continuum events is used to study the detector response, its acceptance, and to validate the analysis techniques. The simulation describes $\Bbar \to D^{**} \ell^- \bar{\nu}_{\ell}$ decays using the ISGW2 model~\cite{ISGW}, and non-resonant $\Bbar
\to D^{(*)} \pi \ell^- \bar{\nu}_{\ell}$ decays using the model of Goity and Roberts~\cite{Goity}.

We select semileptonic $\Bbar \to D^{**}\ell^-\bar{\nu}_{\ell}$ 
decays with $\ell=e, \mu$ in events containing
a fully reconstructed $B$ meson ($B_\mathrm{tag}$), 
 which allows us to constrain the kinematics, reduce the 
 combinatorial background, and determine the charge and flavor 
 of the signal $B$ meson. $D^{**}$ mesons are reconstructed in the $D^{(*)}\pi^{\pm}$
 decay modes and the different $D^{**}$ states are
 identified by a fit to the invariant mass differences $m(D^{(*)}\pi) -
 m(D^{(*)})$.  

We first reconstruct the semileptonic $B$ decay, selecting a lepton with momentum $p^*_{\ell}$ in the  CM  frame larger than 0.6 GeV/$c$. We search for pairs of oppositely-charged tracks that form a vertex and remove those with an invariant mass consistent  with a photon conversion or a $\pi^0$ Dalitz decay.  Candidate $D^0$ mesons that have the correct 
charge correlation with the lepton are reconstructed
in the $K^-\pi^+$, $K^- \pi^+ \pi^0$, $K^- \pi^+ \pi^+ \pi^-$,
$K^0_S \pi^+ \pi^-$, $K^0_S \pi^+ \pi^- \pi^0$, $K^0_S \pi^0$, $K^+ K^-$,
$\pi^+ \pi^-$, and $K^0_S K^0_S$ channels, and $D^+$ mesons in the
$K^- \pi^+ \pi^+$, $K^- \pi^+ \pi^+ \pi^0$, $K^0_S \pi^+$, $K^0_S \pi^+ \pi^0$,
$K^+ K^- \pi^+$, $K^0_S K^+$, and $K^0_S \pi^+ \pi^+ \pi^-$ channels.
In events with multiple $D\ell^-$ combinations, the candidate with the best  $D$-$\ell$ vertex fit is selected.
Candidate $D^*$ mesons are reconstructed by combining a $D$ candidate with a pion or a photon in the $D^{*+} \rightarrow D^0 \pi^+ $, $D^{*+} \rightarrow D^+ \pi^0$, $D^{*0} \rightarrow D^0 \pi^0$, and $D^{*0} \rightarrow D^0 \gamma$ channels. 
In events with multiple $D^{*}\ell^-$ combinations, we choose the candidate with the smallest $\chi^2$ based on the deviations from the nominal values of the $D$ invariant mass and the invariant mass difference between the $D^*$ and the $D$, using the  resolution measured in each mode. 

 We reconstruct $B_\mathrm{tag}$ decays~\cite{BrecoVub} in charmed hadronic modes $\Bbar \rightarrow D Y$, where 
$Y$ represents a collection of hadrons, composed
of $n_1\pi^{\pm}+n_2 K^{\pm}+n_3 K^0_S+n_4\pi^0$, where $n_1+n_2 =1,3,5$, $n_3
\leq 2$, and $n_4 \leq 2$. Using $D^0(D^+)$ and $D^{*0}(D^{*+})$ as seeds for $B^-(\Bzb)$ decays, we reconstruct about 1000 different decay chains.

The kinematic consistency of a $B_\mathrm{tag}$ candidate with a $B$ meson decay is evaluated using two variables: the beam-energy
substituted mass $m_{ES} \equiv \sqrt{s/4-|p^*_B|^2}$, and the energy difference $\Delta E \equiv E^*_B -\sqrt{s}/2$. Here $\sqrt{s}$ is the total CM  energy, and $p^*_B$ and $E^*_B$ denote the momentum and energy of the $B_\mathrm{tag}$ candidate in the CM frame. For correctly identified $B_\mathrm{tag}$ decays, the $m_{ES}$ distribution peaks at the $B$ meson mass, while $\Delta E$ is consistent
with zero.
We select $B_\mathrm{tag}$ candidates in the signal region
defined as 5.27~GeV/$c^2$ $< m_{ES} <$ 5.29~GeV/$c^2$, excluding those with
 daughter particles in common with the charm meson or
the lepton from the semileptonic $B$ decay. In the case of multiple $B_\mathrm{tag}$ candidates in an event, we select the one with the smallest
$|\Delta E|$ value. The $B_\mathrm{tag}$ and the $D^{(*)}\ell$ candidates are required to have the correct charge-flavor correlation. We account for mixing effects in the $\Bzb$ sample as described in Ref.~\cite{BBmixing}. 
Cross-feed effects, $i.e.$, $B^-_\mathrm{tag} (\Bzb_\mathrm{tag})$ candidates erroneously reconstructed as a neutral~(charged) $B$,  are subtracted using estimates from the simulation.

We reconstruct $B^- \to D^{(*)+}\pi^- \ell^- \bar{\nu}_{\ell}$ and $\Bzb \to D^{(*)0}\pi^+ \ell^- \bar{\nu}_{\ell}$ decays starting from the corresponding $B_\mathrm{tag}+D^{(*)} \ell^-$ combinations. We select events with only one additional reconstructed charged track, correctly matched to the $D^{(*)}$ flavor, that has not been used for the reconstruction of the $B_\mathrm{tag}$, the signal $D^{(*)}$, or the lepton. 
$D(D^{*})$ candidates are selected within 2$\sigma$~(1.5-2.5$\sigma$, depending on the $D^*$ decay mode) of the $D$ mass ($D^{*}-D$ mass difference), where the resolution $\sigma$ is typically around 8~(1-7) MeV$/c^{2}$. 
For the $\Bzb \rightarrow D^{(*)0}\pi^+ \ell^- \bar{\nu}_{\ell}$ decay,
we additionally require the invariant mass difference $m(D^0\pi^+)-m(D^0)$ to be greater than 0.18 GeV/$c^2$ to veto $\Bzb \rightarrow D^{*+} \ell^- \bar{\nu}_{\ell}$ events.

Semileptonic $\Bbar \rightarrow D^{**}\ell^- \bar{\nu}_{\ell}$ decays are identified by the missing mass squared in the event, $m^2_\mathrm{miss} = \left[p(\Upsilon(4S)) -p(B_\mathrm{tag}) - p(D^{(*)}\pi) - p(\ell)\right]^2$, defined in terms of the particle four-momenta. For correctly reconstructed signal events, the only missing particle is the neutrino, and $m^2_\mathrm{miss}$ peaks at zero. Other $B$ semileptonic decays, where one particle is not reconstructed (feed-down) or is erroneously added to the charm candidate (feed-up), exhibit higher or lower values in $m^2_\mathrm{miss}$~\cite{babar-3}. In feed-down cases where both  a $D$ and a $D^*$ candidate have been reconstructed, we keep only the latter candidate.

\begin{table}[!t]
\caption{$m^2_\mathrm{miss}$ selection criteria.}
\centering
\begin{tabular}{ll}
\hline
\hline
Mode & Selection Criteria \\
\hline
$B^- \to D^{*+}\pi^- \ell^- \bar{\nu}_{\ell}$ &  $-0.25 < m^2_\mathrm{miss} < 0.25$ GeV$^2/c^4$\\
$B^- \to D^{+}\pi^- \ell^- \bar{\nu}_{\ell}$ &  $-0.25 < m^2_\mathrm{miss} < 0.8$ GeV$^2/c^4$\\
$\Bzb \to D^{*0}\pi^+ \ell^- \bar{\nu}_{\ell}$ &  $-0.2 < m^2_\mathrm{miss} < 0.35$ GeV$^2/c^4$\\
$\Bzb \to D^{0}\pi^+ \ell^- \bar{\nu}_{\ell}$ &  $-0.15 < m^2_\mathrm{miss} < 0.85$ GeV$^2/c^4$\\
\hline
\hline
\end{tabular}
\label{tab:MMcuts}
\end{table}

The $m^2_\mathrm{miss}$ selection criteria are listed in Table \ref{tab:MMcuts}. The $m^2_\mathrm{miss}$ region between 0.2 and 1 GeV$^{2}/c^{4}$ for $\Bbar \rightarrow D \pi\ell^- \bar{\nu}_{\ell}$ events is dominated by feed-down from $\Bbar \to D^{**} (\to D^* \pi) \ell^- \bar{\nu}_{\ell}$ semileptonic decays where the soft pion from the $D^*$ decay
is not reconstructed. In order to retain these events we apply an asymmetric cut on $m^2_\mathrm{miss}$ for these modes. 

The signal yields for the $\Bbar \to D^{**}\ell^- \bar{\nu}_{\ell}$ decays are extracted through a simultaneous unbinned maximum likelihood fit to the four $m(D^{(*)}\pi) - m(D^{(*)})$ distributions. 
With the current statistics, validation studies on MC samples show that our sensitivity to non-resonant $\Bbar \to D^{(*)}\pi \ell^- \bar{\nu}_{\ell}$ decays is limited. Including hypotheses for these components results in a fitted contribution that is consistent with zero. Thus we assume that these non-resonant contributions are negligible.
The probability that $\Bbar \to D^{**} (\to D^* \pi) \ell^- \bar{\nu}_{\ell}$ decays are reconstructed as $\Bbar \to D^{**} (\to D \pi) \ell^- \bar{\nu}_{\ell}$ is determined with the MC simulation to be 26\%(59\%) for the $B^-$($\Bzb$) sample and held fixed in the fit.

\begin{figure}[!ht]
\includegraphics[width=0.5\textwidth,totalheight=0.62\textheight]{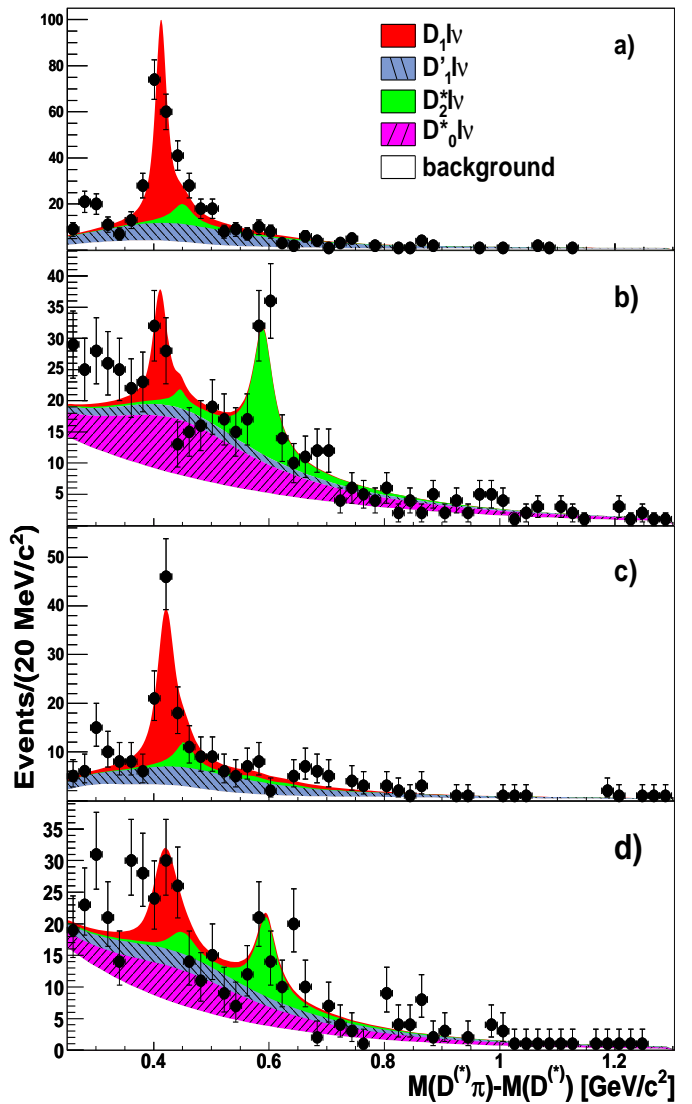}
\label{fig:Fit}
\caption{(Color online) Fit to the $m(D^{(*)}\pi)-m(D^{(*)})$ distribution for a) $B^- \to D^{*+}\pi^- \ell^- \bar{\nu}_{\ell}$, b) $B^- \to D^+ \pi^- \ell^- \bar{\nu}_{\ell}$, c) $\Bzb \to D^{*0} \pi^+ \ell^- \bar{\nu}_{\ell}$, and d) $\Bzb \to D^0 \pi^+ \ell^- \bar{\nu}_{\ell}$: the data (points with error bars) are compared to the results of the overall fit (sum of the solid distributions). The PDFs for the different fit components are stacked and shown in different colors.}
\end{figure}

The Probability Density Functions (PDFs) for the $D^{**}$ signal components are determined using MC $\Bbar \to D^{**}\ell^- \bar{\nu}_{\ell}$ signal events. A convolution of a Breit-Wigner function with a Gaussian, whose resolution is determined from the simulation, is used to model the $D^{**}$ resonances. The $D^{**}$ masses and widths are fixed to measured values~\cite{pdg}. We rely on the MC prediction for the shape of the combinatorial and continuum background. A non-parametric KEYS function~\cite{keys} is used to model this component for the $D^* \pi \ell^- \bar{\nu}_{\ell}$ sample, while for the $D \pi \ell^- \bar{\nu}_{\ell}$ sample we use the convolution of an exponential with a Gaussian to model the tail from virtual $D^*$ mesons. 
The combinatorial and continuum background yields are estimated from data. We fit the hadronic $B_\mathrm{tag}$ $m_{ES}$ distributions for $\Bbar \to D^{**}\ell^- \bar{\nu}_{\ell}$ events as described in~\cite{babar-3}, and we obtain the number of background events from the integral of the background function in the $m_{ES}$ signal region.  

Table \ref{tab:results} summarizes the results from two fits: one in which we fit the charged and neutral $B$ samples  separately, and one in which we impose the isospin constraints ${\cal B} (B^- \to D^{**}\ell^- \bar{\nu}_{\ell})/{\cal B} (\Bzb \to D^{**}\ell^- \bar{\nu}_{\ell}) = \tau_{B^-}/\tau_{\Bzb}$. The latter fit yields a significance greater than 6
 standard deviations for all four $D^{**}$ states including systematic uncertainties. The results of this fit are shown in Fig.\ 1\ref{fig:Fit}.

The $D_2^*$ contributes to both the $D\pi$ and the $D^*\pi$ samples. In the nominal fit we fix the ratio ${\cal B} (D^*_2 \to D\pi)/{\cal B} (D^*_2 \to D^*\pi)$ to $2.2$~\cite{pdg}. When we allow this ratio to float we obtain $1.9 \pm 0.6$. 

To reduce systematic uncertainties we measure the ratios of the ${\cal B} (\Bbar \rightarrow D^{**} \ell^- \bar{\nu}_{\ell})$ branching fractions to the inclusive $\Bzb$ and $B^-$ semileptonic branching fractions.  
A sample of $\Bbar \to X \ell^- \bar{\nu}_{\ell}$ events is selected by identifying a charged lepton with $p^*_{\ell}>0.6$ GeV/$c$  and the correct charge correlation with the $B_\mathrm{tag}$ candidate. In the case of multiple $B_\mathrm{tag}$ candidates in an event, we select the one reconstructed in the decay channel with the highest purity, defined as the fraction of signal events in the $m_{ES}$ signal region. 
Background components that peak in the $m_{ES}$ signal region include cascade $B$ meson decays ($i.e.$, the lepton does not come directly from the $B$) and hadronic decays, and are subtracted using the corresponding MC predictions.  

\begin{table*}[!htb]
\caption{Results from the fits to data: the $\Bbar \to D^{**} \ell^- \bar{\nu}_{\ell}$ signal yield, the corresponding reconstruction efficiency, the product of branching fractions, where the first error is statistical and the second systematic. For the $\Bbar \to D_2^* \ell^- \bar{\nu}_{\ell}$ decay, we report yields and product of branching fractions for the $D_2^* \to D\pi$ decay mode. For the isospin-constrained results (last two columns), the $B^-$ branching fraction products are reported. The statistical significances, $S_\mathrm{stat}$, are obtained by computing the difference in the log
likelihood between the nominal fit and the fit in which we fix the different signal components to 0. The significances including the systematic uncertainty, $S_\mathrm{tot}$, are obtained by rescaling the statistical significances by $\sigma_\mathrm{stat}/\sqrt{\sigma^2_\mathrm{stat}+\sigma^2_\mathrm{syst}}$.}
\begin{tabular}{lrcclcl}
\hline
\hline
Decay Mode & Yield & $\epsilon_\mathrm{sig} (\times 10^{-4})$ & ${\cal B}$ ($\Bbar \to D^{**} \ell^- \bar{\nu}_{\ell}$ ) $\times$ ${\cal B} (D^{**} \to D^{(*)} \pi^{\pm})$ \% & $S_\mathrm{tot} (S_\mathrm{stat})$ & ${\cal B}$ \% & $S_\mathrm{tot} (S_\mathrm{stat})$\\
\hline
$B^- \to D^{0}_1 \ell^- \bar{\nu}_{\ell}$ & $165 \pm 18$ & 1.24 & $0.29 \pm 0.03 \pm 0.03$ & 9.9~(12.7) & $0.29 \pm 0.03 \pm 0.03$ &10.7~(15.2) \\
$B^- \to D^{*0}_2 \ell^- \bar{\nu}_{\ell}$ & $97 \pm 16$ & 1.44 & $0.15 \pm 0.02 \pm 0.01$ & 6.3~(7.3) & $0.12 \pm 0.02 \pm 0.01$  &6.0~(7.4)\\
$B^- \to D^{'0}_1 \ell^- \bar{\nu}_{\ell}$ & $142 \pm 21$ & 1.13 & $0.27 \pm 0.04 \pm 0.05$ &5.4~(8.0) & $0.30 \pm 0.03 \pm 0.04$ &6.4~(10.0)\\
$B^- \to D^{*0}_0 \ell^- \bar{\nu}_{\ell}$ & $137 \pm 26$ & 1.15 & $0.26 \pm 0.05 \pm 0.04$  &4.5~(5.8) & $0.32 \pm 0.04 \pm 0.04$ &6.1~(8.3)\\
$\Bzb \to D^{+}_1 \ell^- \bar{\nu}_{\ell}$ & $88 \pm 14$ & 0.70 & $0.27 \pm 0.04 \pm 0.03$ &7.0~(8.4) & & \\
$\Bzb \to D^{*+}_2 \ell^- \bar{\nu}_{\ell}$ & $29 \pm 13$ & 0.91 & $0.07 \pm 0.03 \pm 0.01$ ($< 0.11$ @90\% CL) &2.3~(2.5) & &\\
$\Bzb \to D^{'+}_1 \ell^- \bar{\nu}_{\ell}$ & $86 \pm 18$ & 0.60 & $0.31 \pm 0.07 \pm 0.05$ &4.6~(5.8) & &\\
$\Bzb \to D^{*+}_0 \ell^- \bar{\nu}_{\ell}$ & $142 \pm 26$ & 0.70 & $0.44 \pm 0.08 \pm 0.06$ &4.7~(6.0) & &\\
\hline
\hline
\end{tabular}
\label{tab:results}
\end{table*}

The total yield for the inclusive $\Bbar \to X \ell^- \bar{\nu}_{\ell}$ decays is obtained from a maximum likelihood fit to the $m_{ES}$ distribution of the $B_\mathrm{tag}$ candidates, as described in ~\cite{babar-3}. The fit yields 198,897 $\pm$ 1,578 events for the 
$B^- \to X \ell^- \bar{\nu}_{\ell}$ sample and 120,168 $\pm$ 1,036 events for the $\Bzb \to X \ell^- \bar{\nu}_{\ell}$ sample. 

The ratios ${\cal B}(\Bbar \to D^{**} \ell^- \bar{\nu}_{\ell})/{\cal B}(\Bbar \to X \ell^- \bar{\nu}_{\ell})= (N_\mathrm{sig}/\epsilon_\mathrm{sig})\cdot (\epsilon_\mathrm{sl}/N_\mathrm{sl})$ are obtained by correcting the signal yields for the reconstruction efficiencies (estimated  from \BB\ MC events). Here, $N_\mathrm{sig}$ is the number of 
$\Bbar \to D^{**} \ell^- \bar{\nu}_{\ell}$ signal events, reported in  Table~\ref{tab:results} together
 with the corresponding reconstruction efficiencies
$\epsilon_\mathrm{sig}$, $N_\mathrm{sl}$ is the $\Bbar \to X \ell^- 
\bar{\nu}_{\ell}$ signal yield, and $\epsilon_\mathrm{sl}$ 
is the corresponding reconstruction efficiency 
including the $B_\mathrm{tag}$ reconstruction, 
equal to 0.39\% and 0.25\% for the $B^- \to X \ell^- \bar{\nu}_{\ell}$ 
and $\Bzb \to X \ell^- \bar{\nu}_{\ell}$ decays, respectively. The absolute branching fractions ${\cal B} (\Bbar \rightarrow D^{**} \ell^- \bar{\nu}_{\ell})$ are then determined using the semileptonic branching fraction ${\cal B}(\Bbar \to X \ell^- \bar{\nu}_{\ell})= ( 10.78 \pm 0.18)\%$ and the ratio of the $\Bzb$ and the $B^-$ lifetimes $\tau_{B^-}/\tau_{\Bzb} = 1.071 \pm 0.009$~\cite{pdg}.

Numerous sources of systematic uncertainties have been investigated.
The largest uncertainty is due to the determination of the $\Bbar \to D^{**}\ell^- \bar{\nu}_{\ell}$ signal yields  (resulting in 5.5-17.0\% relative systematic uncertainty depending on the $D^{**}$ state). This uncertainty is estimated using ensembles of fits to the data in which the input parameters are varied within the known uncertainties in the PDF parameterization (0.2-8.7\%), the shape and yield of the combinatorial and continuum background (0.2-10.4\%), the modeling of the broad $D^{**}$ states (4.5-13.8\%), and the $D^{*}$ feed-down rate (0.5-4.0\%). 
We check that the combinatorial and continuum background shape is well reproduced by the simulation by verifying that the MC samples of right-sign and wrong-sign $D^{(*)}\pi$ combinations have similar shapes, and that the wrong-sign distribution in the data agrees well with that in the simulation. 
We observe an excess of events in the low invariant mass difference region in
the four samples that is not accounted for by the background PDF. 
We study $\Bbar \to D^{(*)}n\pi \ell^-\bar{\nu}_{\ell}$ ($n>1$) decays, not included in our standard MC simulation, as a possible source of this excess. We use different MC models for these decays, and find that they do not account for all the observed excess. We evaluate a corresponding  
systematic uncertainty (0.1-3.2\%), included in the yield uncertainty above. 
The uncertainties due to the detector simulation are determined by varying, within bounds given by data control samples, the charged track reconstruction efficiency (1.3-2.0\%), the photon reconstruction efficiency (0.2-4.8\%), the lepton identification efficiency (1.2-1.6\%),  and the reconstruction efficiency for low momentum charged (1.2\%) and neutral pions (1.3\%).  
We use an HQET model~\cite{LLSW} to test the model dependence of the $\Bbar \to D^{**} \ell^- \bar{\nu}_{\ell}$ simulation (0.8-2.5\%). We include the uncertainty on the branching fractions of the reconstructed $D$ and $D^{*}$ modes (3.0-4.5\%), and on the absolute branching fraction ${\cal B} (\Bbar \to X \ell^- \bar{\nu}_{\ell})$ used for the normalization (1.9\%).
We also include a systematic uncertainty due to differences in the efficiency of the $B_\mathrm{tag}$ selection in the exclusive selection of $\Bbar \to D^{**} \ell^- \bar{\nu}_{\ell}$ decays and the inclusive $\Bbar \to X \ell^- \bar{\nu}_{\ell}$ reconstruction (4.0-5.6\%). 

In conclusion, we report the simultaneous observation of $\Bbar \to D^{**}\ell^-\bar{\nu}_{\ell}$ decays into the four $D^{**}$ states predicted by HQS. The measured branching fractions are reported in Table \ref{tab:results}.  We find results consistent with Ref.~\cite{babar-3} for the sum of the different $D^{**}$ branching fractions. The rate for the $D^{**}$ narrow states is in good agreement with recent measurements~\cite{D0}; the one for the broad states is in agreement with DELPHI~\cite{delphi2005} but does not agree with the $D'_1$ limit of Belle~\cite{belle}. The rate for the 
broad states is found to be large.  If these broad states are indeed due 
to $\Bbar \to D'_1 \ell^- \bar{\nu}_{\ell}$ and $\Bbar \to D^*_0 \ell^- \bar{\nu}_{\ell}$ decays, this is in conflict with the 
expectations from QCD sum rules.

\input acknow_PRL.tex

\end{document}

%% file: authors_jun2008.tex
%
\author{B.~Aubert}
\author{M.~Bona}
\author{Y.~Karyotakis}
\author{J.~P.~Lees}
\author{V.~Poireau}
\author{E.~Prencipe}
\author{X.~Prudent}
\author{V.~Tisserand}
\affiliation{Laboratoire de Physique des Particules, IN2P3/CNRS et Universit\'e de Savoie, F-74941 Annecy-Le-Vieux, France }
\author{J.~Garra~Tico}
\author{E.~Grauges}
\affiliation{Universitat de Barcelona, Facultat de Fisica, Departament ECM, E-08028 Barcelona, Spain }
% ...
\collaboration{The \babar\ Collaboration}
\noaffiliation

%% file: acknow_PRL.tex
We are grateful for the excellent luminosity and machine conditions
provided by our \pep2\ colleagues, 
and for the substantial dedicated effort from
the computing organizations that support \babar.
The collaborating institutions wish to thank 
SLAC for its support and kind hospitality. 
This work is supported by
DOE
and NSF (USA),
NSERC (Canada),
CEA and
CNRS-IN2P3
(France),
BMBF and DFG
(Germany),
INFN (Italy),
FOM (The Netherlands),
NFR (Norway),
MIST (Russia),
MEC (Spain), and
STFC (United Kingdom). 
Individuals have received support from the
Marie Curie EIF (European Union) and
the A.~P.~Sloan Foundation.

%% file: paper.bbl
\begin{thebibliography}{99}
\bibitem{ell}
The charge conjugate state is always implied unless stated otherwise.
\end{thebibliography}
